\documentclass[twocolumn,showpacs,preprintnumbers,amsmath,amssymb]{revtex4}

\usepackage{graphicx}
\usepackage{epsfig}
\usepackage{bm}% bold math

\usepackage{amssymb}

\begin{document}
%\center{\it  Version 1. Preprint (2004)   }

\title{The Geometrical Structure of  Disordered Sphere Packings }

\author{ T. Aste, M. Saadatfar and T.J. Senden}
\affiliation{ Department of Applied Mathematics, Research School of Physical Sciences and Engineering,  The Australian National University, 0200 Australia.}

%\author{}
\date{ \today }% It is always \today, today,

\begin{abstract}
% Text of abstract
The three dimensional structure of large packings of monosized spheres with volume fractions ranging between 0.58 and 0.64 has been studied with X-ray Computed Tomography.
We search for signatures of organization, we classify local arrangements and we explore the effects of local geometrical constrains on the global packing.
This study is the largest and the most accurate empirical analysis of disordered packings at the grain-scale to date with over 140,000 sphere coordinates mapped.
We discuss topological and geometrical ways to characterize and classify these systems, and discuss implications that local geometry can have on the mechanisms of formation of these amorphous structures. 
\end{abstract}


% main text

\maketitle

\section{Introduction}
\label{s.I}

When balls are poured into a container they arrange themselves in a disorderly fashion with no obvious symmetries or repetitive patterns.
However, disorder does not mean randomness.
Indeed such systems are locally highly structured in a hierarchical organization which tries to achieve the goal of maximal compaction under the unavoidable geometrical constraints of non-interpenetration, satisfying simultaneously the condition of force and torque balance on each ball.
This leads to very complex structures which show signs of organization but nevertheless have so far eluded all efforts for a simple and clear classification.
In order to fully classify the state of a disordered system, such as a granular packing at rest, the exhaustive details about the exact position, orientation and shape of each grain is, in principle, needed.
However, part of such information is at best redundant or even irrelevant and several degenerate states with different microscopic realizations can share the same macroscopic properties. 
To determine which are the accessible configurations at the local level,  and to understand  which are the possible combinations which generate the global packing is  of fundamental importance.
Indeed, finding measures for the local and global, hierarchical organization is the essential starting point towards the understanding of the basic mechanisms which form these structures.
It is also an essential step in the development of technologies which enable us to control and `tune' the structure of amorphous materials.

Until now the empirical investigation of the geometrical structure of these systems have been limited by the very little availability of accurate experimental data.
Indeed, after the seminal works of Bernal, Mason and Scott \cite{Bernal60,Scott62,Mason68}, it has been only very recently that the use of tomography has allowed to investigate three dimensional structure  from the grain level up to the whole packing.

\begin{figure}
\begin{center}
\resizebox{0.30\textwidth}{!}{%
%\begin{tabular}[c]{cc}
\includegraphics{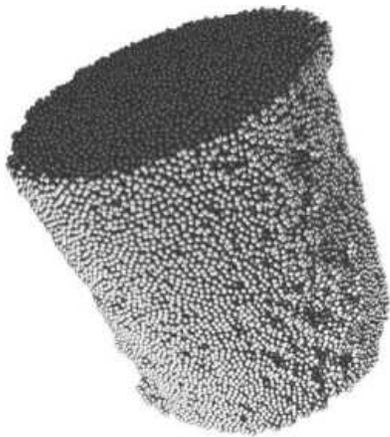} 
%&
%\mbox{\epsfig{file=distance.eps,width=5.cm,angle=0}}
%\end{tabular}
}
\end{center}
\caption{ A reconstruction of a packing of $\sim 150,000$ spheres in a cylindrical container  (sample C).   }
\label{f.pack}
\end{figure}

The first work which uses tomographic techniques devoted to the investigation of granular packing is by  Seidler et al.   in 2000  \cite{Seidler00}.
Other two works by Sederman et al. \cite{Sederman01} and  Richard et al. \cite{Richard03} followed respectively in 2001 and 2003.
Confocal microscopy techniques have been also used to reconstruct 3D images of a dense packing emulsion of oil droplets \cite{Brujic03} and to count contacts in glass beads \cite{Kohonen04}.
However, all these works concern rather small sample sizes and focus the analysis  on few particular topics.
In this paper we present an empirical investigation by means of X-ray Computed Tomography on very large disordered packings of monosized spheres with packing densities  ranging from 0.58 to 0.64. 
(The packing density is the fraction of volume occupied by the balls divided by the total volume, and it is often called  `volume fraction' in the literature).
This study is the largest and the most accurate empirical analysis of disordered packings at the grain-scale ever performed.
A packing realization is shown in Fig.\ref{f.pack}.
Preliminary results were presented in ref. \cite{AsteKioloa}.
Here we perform a more extensive and complete investigation using an improved algorithm to calculate the positions of the spheres. 
Additional material including the sphere coordinations of some of the samples can be found in  \cite{WebGrain}.

In  order to help the reader this paper is organized in several sections each addressing  different aspects.
Each section has been designed to be as self-containing as possible.
Cross references among sections guide the reader who may wish to focus on specific topics. 

In Section \ref{s.II} the experimental apparatus and the relevant methodology to extract geometrical information from the tomography data are described. 

In Section \ref{s.III} the number of neighbors for each sphere in the packing is studied and a new tool to deconvolute  the contribution of touching neighbors  from the contribution from near neighbors is introduced. 
Implications on mechanical equilibrium are also discussed.

In Section \ref{s.IV} the hierarchical structure of the contact network is analyzed in terms of a shell map \cite{ASBORI}. 

The local symmetries are explored in Section \ref{s.V} by using spherical harmonics decomposition \cite{Steinhardt83}. 

In Section \ref{s.VI}, the compactness of the average local packing is discussed and compared with crystalline packings. 

We analyze and discuss the results for the radial distribution function in Section \ref{s.VII} . 

The density fluctuations at sample level and at grain level are investigated in Section \ref{s.VIII}. 
The implication that  sample geometry can have on the dynamical formation of these systems is discussed in Section \ref{s.IX}. 

A conclusion summarizes the main results and perspectives.

%\vspace*{3cm}

\section{Experimental Apparatus and Methodology}
\label{s.2}
\label{s.II}
The empirical studies reported in this paper concern the analysis of 6 samples made of mono-sized acrylic beads packed in a cylindrical container with an inner diameter of $ 55\; mm$ and filled  to a height of $\sim 75\; mm$. 
In particular we have:
\begin{itemize}
\item Two large samples containing $\sim 150,000$ beads with diameters $d =
1.000\;  mm $ and polydispersity within $0.05 \; mm$;
\item Four smaller samples containing $\sim 35,000$ beads with diameters $d = 1.59 \; mm $ and polydispersity within $0.05 \; mm$.
\end{itemize}

An independent estimation of the polydispersity for the small spheres was performed by weighing 200 beads and  computing their standard deviation.
The estimated value for the relative statistical error on the sphere diameters is 1.5\%.

\subsection{\bf Sample Preparation}
The 6 samples (named hereafter with labels A-F) have been prepared at different packing densities $\rho$ ranging between 0.586 (sample A) to 0.640 (sample F).
Table~\ref{t.1} reports in the second column all the sample densities.  
The two packings at lower densities (A, B respectively with $\rho \sim 0.586-0.596$) were obtained by placing a stick in the middle of the container before pouring the beads into it and then slowly removing the stick \cite{ppp}. 
Sample C ($\rho = 0.619$) was obtained by gently and slowly pouring the spheres into the container.
Whereas, the sample D ($\rho = 0.626$) was obtained by a faster pouring. 
Higher densities, up to $\rho \sim 0.63$, were achieved by gently tapping the container walls.  
The densest sample at $\rho = 0.64$ was obtained by a combined action of gently tapping and compression from above (with the upper surface left unconfined at the end of the preparation). 
To reduce boundary effects, the inside of the cylinder has been roughened by randomly gluing spheres to the internal surfaces.

\subsection{\bf XCT Imaging} 

A X-ray Computed Tomography apparatus (see Sakellariou \textit{et al}.  \cite{Sakellariou04}) is used to measure the density maps of the samples.  
The two large samples (A, C) were analysed by acquiring data sets of $2000^3$ voxels with a spatial resolution of
$0.03 \;mm$; whereas the four smaller samples (B, D, E, F) were analysed by acquiring data sets of $1000^3$ voxels with a spatial resolution of $0.06 \; mm$.
After segmentation (see Sheppard \textit{et al} \cite{Sheppard04}) the sample data sets are reduced to three-dimensional binary images, representing two distinct phases, one associated with the spheres and the other with air space.
The effective spatial resolution of this technique is limited by the finite size of the X-ray source, surface scattering of the low energy X-ray and intrinsic blurring from reconstruction.
From a careful analysis of the reconstructed samples we observed that the combination of all these factors generates some fuzziness in a region between one and two voxels around the sphere surfaces.
 
 %\vspace*{3cm}
\begin{figure}
\begin{center}
\resizebox{0.40\textwidth}{!}{%
\includegraphics{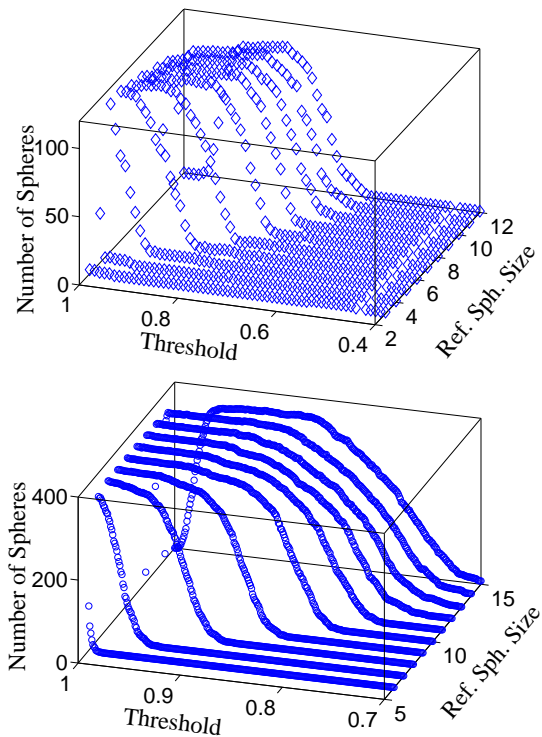} 
}
\end{center}
\caption{ 
Number of detected spheres in an internal region of samples A (top) and D (bottom) reported as function of the reference sphere radius and the threshold. 
Threshold is expressed in relative units with respect to the value of the highest intensity peak in that region. 
}
\label{f.tresh}
\end{figure}

\subsection{\bf Sphere Centres} 
In order to proceed with the analysis of the geometrical and statistical properties, the position of all sphere centres are calculated from three dimensional binary images.  
Our approach is to find the sphere centres by moving a reference sphere ($S$) throughout the packing ($P$) and measuring the local overlap between $S$ and $P$. 
This corresponds to a 3-dimensional convolution: $P*S$. 
The regions with larger overlaps are the ones around the centers of the spheres in the packing.
In order to isolate these regions we apply a threshold on the intensity map resulting from the convolution.
The centres of mass of such regions are good estimations for the positions of the packed spheres.
This method is made numerically highly efficient by applying the convolution theorem which allows to transform the convolution into a product in Fourier space: ${\mathcal F}[P*S]= {\mathcal F}[P]{\mathcal F}[S]$, where ${\mathcal F}$ represents the (fast)Fourier Transform. 
The algorithm proceeds in 4 steps: {\bf 1)} fast Fourier transform of the binary image (${\mathcal F}[P]$); {\bf 2)} transform the digitised map of the reference sphere (${\mathcal F}[S]$); {\bf 3)} perform the direct product between these two; {\bf 4)} inverse-transform of the product:
${\mathcal F}^{-1}[{\mathcal F}[P]{\mathcal F}[S]]= P*S $.  
The result is an intensity map of the overlap between the reference sphere and the bead pack, where the voxels closer to the sphere centres have a highest intensity.  
A threshold on the intensity map locates the groups of voxels surrounding the sphere centers which become isolated clusters.
The sphere centers are calculated as the centre of mass (intensity) of these clusters.

The precision on the estimation of the sphere-center positions can be evaluated considering that the spatial resolution is within one and two voxels.
Therefore the precision on the centre of mass of a cluster of $\nu$ voxels must be within $2/\nu$.
%However, we know that a layer of $\Delta$ voxels on the boundary of each grain is affected by  surface scattering effects  (...) reducing therefore the effective size of the cluster to the internal part only and consequently reducing the precision to $\sim [\nu - (48 \pi^2 \nu)^{1/3} \Delta]^{-1}$.
In the procedure to locate the sphere-centers one has two dajusTable~parameters: the reference sphere size and the  threshold.
We have searched for the optimal choice of the parameters by varying these two quantities and computing the resulting number of spheres detected in a given portion of the sample.
In Fig.\ref{f.tresh} the number of detected spheres is reported in function of the reference sphere size and the threshold. 
When the reference sphere is too small only one  cluster is detected independently on the threshold.
Similarly when the threshold is too small, peaks cannot be isolated and one spanning cluster is also observed.
On the other hand there is a rather large region of the two parameters where the same amount of spheres is detected.
We know that  precision increases with the size of the clusters on which the center of mass is computed.
Therefore the best choice of parameters is the one which leads to the largest clusters.
This requires the smallest possible threshold and reference sphere compatibly with the correct detection of each sphere in the system.
We chose the threshold at 0.95 (samples  A, C, E ) and at 0.93 (samples B, D, F), and we fixed the reference  sphere radius at  13  (sample A, C), 11 (sample E) and 10 (samples B, D, F).
Obtaining typical cluster sizes of $\sim 80$ voxels (samples B, D, E, F) and $\sim 400$ (samples A, C). 
This implies precisions on the sphere centers respectively  within 3\% and 0.5 \% of the voxel sizes \footnote{Some of the datasets with the sphere coordinates, and other material, are available at: http://wwwrsphysse.anu.edu.au/granularmatter/}.

\subsection{\bf Central Region}
In order to reduce boundary effects, all the analysis reported hereafter have been performed over a central region ($\mathbf G$) at 4 sphere-diameters away from the sample boundaries.
Note that spheres outside $\mathbf G$ are considered when computing the neighbouring environment of spheres in $\mathbf G$.
In Table~\ref{t.1} the number of spheres in this region ($N_G$) is reported for each sample.

\begin{table*}
\begin{tabular}[c]{cccccccccc}
\hline
 & density & N& $N_G$ 	&  $n_c$  & $n_t(1)$ 		& $n_t(1.02)$ 		& $n_t(1.05)$ 	& $n_t(1.1)$ & $ \xi$  \\
\hline
A & $0.586  \pm 0.005$ & 102,897  & 54,719 & {\bf 5.81} & 3.0&  5.5 & 6.7 & 7.5 &   0.014\\
B  & $0.596 \pm 0.006$ & 34,016    & 15,013 & {\bf 5.91} & 2.9 &  5.9 &  6.8 & 7.7 &   0.011\\
C & $0.619\pm 0.005$ & 142,919  & 91,984  & {\bf 6.77} & 3.5 &  6.4&  7.5 & 8.4  &   0.013\\
D & $0.626 \pm 0.008$ & 35,511    & 15,725  & {\bf 6.78} & 3.3 &  6.0 &  7.5 & 8.4  &  0.017\\
E  & $0.630 \pm 0.01$   & 35,881    & 15,852  & {\bf 6.95}  & 3.4 &  6.3 &  7.6 & 8.6  &   0.016\\
F  & $0.640 \pm 0.005$ & 36,461    & 16,247  & {\bf 6.97} & 3.3 &  6.9 &  7.9 & 8.9 &   0.011\\
\hline
\end{tabular} 
\caption{ 
\label{t.1} Sample density and their interval of variations ($\pm$) within each sample;  number of spheres in the sample ($N$); number of spheres in the central region ($N_G$); estimated average number of neighbors in contact ($n_c$), average number of neighbors at given radial distance ($n_t(r)$ with $r=1,1.02,1.05,1.1$ diameters). Standard deviation ($\xi$) calculated from the probability distribution for radial distances smaller than $d$ between pair of sphere centers.
}
\end{table*}

%%%%%%%%%%%%%%%%%%%%%%%%%%%%%%%%%%%%%
%%%%%%%%%%%%%%%%%%%%%%%%%%%%%%%%%%%%
\begin{figure}
\begin{center}
\resizebox{0.50\textwidth}{!}{%
%\begin{tabular}{r}
%\includegraphics{xfigNt} 
\includegraphics{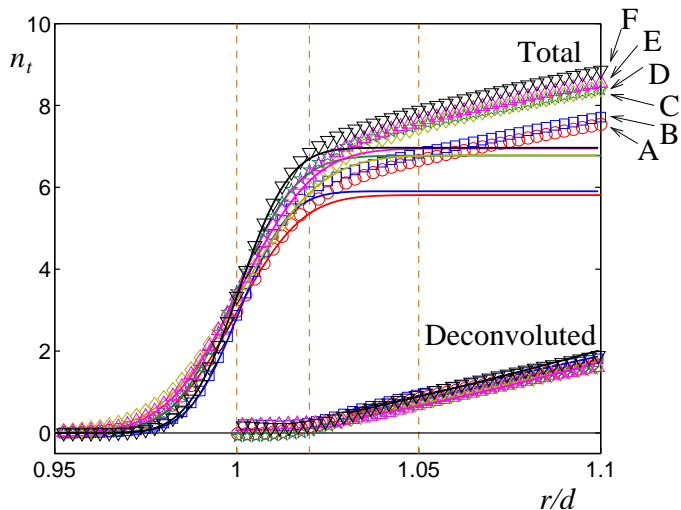}
%\end{tabular}
}
\end{center}
%\vspace{-0.4cm}
\caption{ 
\label{f.c_cum}
({\it symbols}) Behavior for the average number of sphere centers within a radial distance $r$.
({\it lines}) Complementary error function $n_t(r)^{fit}$ (Eq.\ref{ncfit}) normalized by best-fitting the agreement with the data in the region $r<d$.
The averages ($d$) and the standard deviations ($\xi$) are calculated from the probability distribution for radial distances smaller than $d$ between pair of sphere centers.
The re-normalized complementary error function fits well the data for $r<d$. 
After this value near-neighbors not in contact start to contribute significantly to  $n_t(r)$ and the two behaviors split.
The `deconvoluted'  plots show the difference between $n_t(r)$ and the fit with $n_t(r)^{fit}$.
 }
\end{figure}
%%%%%%%%%%%%%%%%%%%%%%%%%%%%%%%%%%%%%
%%%%%%%%%%%%%%%%%%%%%%%%%%%%%%%%%%%%%

%%%%%%%%%%%%%%%%%%%%%%%%%%%%%%%%%%%%%
%%%%%%%%%%%%%%%%%%%%%%%%%%%%%%%%%%%%
\begin{figure}
\begin{center}
\resizebox{0.50\textwidth}{!}{%
%\begin{tabular}{r}
%\includegraphics{xfigNt} 
\includegraphics{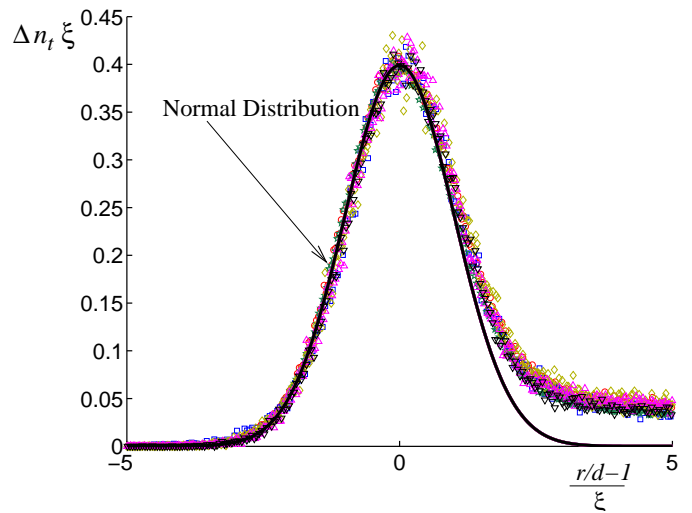}
%\end{tabular}
}
\end{center}
%\vspace{-0.4cm}
\caption{ 
\label{f.gaussCollapse}
 In the region $r<d$ the distribution of radial distances among neighboring spheres are very well mimicked by a Normal distribution ($\Delta n_t(r)$ being the average number of spheres at radial distance between $r-\Delta r$ and $r+\Delta r$) .
All data for all the 6 samples collapse into a single behavior when $\Delta n_t(r) \xi$ is plotted versus $(r/d-1)/\xi$.
The line is the (collapsed) fit of the data in the region $r<d$ (left-hand side) with the Normal distribution. 
This is the same fit used in the deconvolution shown in Fig.\ref{f.c_cum}, which allows to estimate the average number of spheres in contact.
 }
\end{figure}
%%%%%%%%%%%%%%%%%%%%%%%%%%%%%%%%%%%%%
%%%%%%%%%%%%%%%%%%%%%%%%%%%%%%%%%%%%%

%%%%%%%%%%%%%%%%%%%%%%%%%%%%%%%%%%%%%
%%%%%%%%%%%%%%%%%%%%%%%%%%%%%%%%%%%%
\begin{figure}
\begin{center}
\resizebox{0.40\textwidth}{!}{%
%\begin{tabular}{r}
\includegraphics{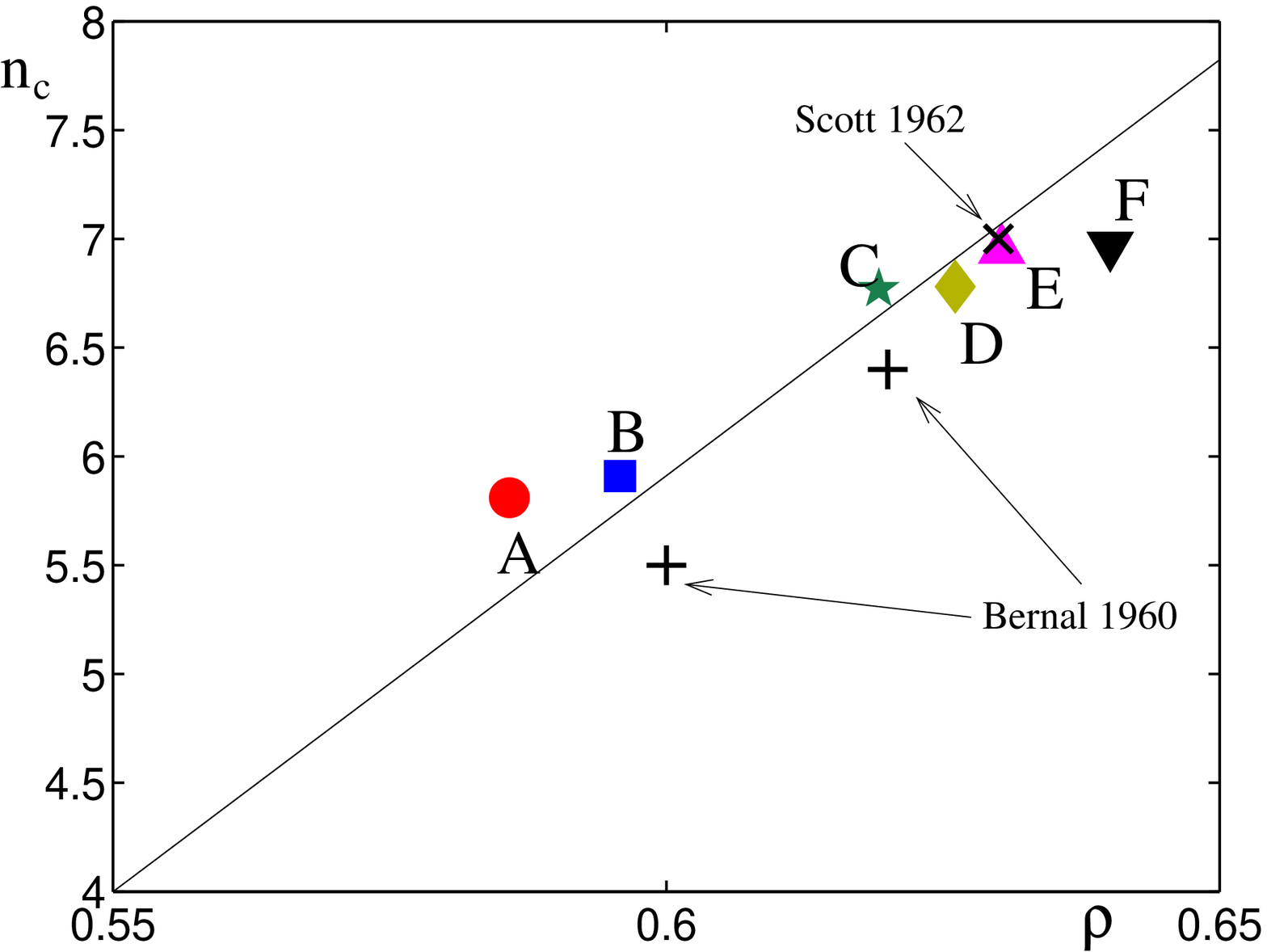} 
%\end{tabular}
}
\end{center}
%\vspace{-0.4cm}
\caption{ 
\label{f.Nc_rho}
Number of neighbors in contact vs. sample density.
The filled symbols corresponds to the samples investigated in the present work.
The two symbols `$+$' are the values from \cite{Bernal60} whereas the `$\times$' is from \cite{Scott62,Mason68}.
The line is the best-fitting of the A-F data with a linear trend constrained to  pass through $n_c = 4$ at $\rho = 0.55$.
}
\end{figure}
%%%%%%%%%%%%%%%%%%%%%%%%%%%%%%%%%%%%%
%%%%%%%%%%%%%%%%%%%%%%%%%%%%%%%%%%%%%

\section{Number of Neighbors in Contact}
\label{s.NNC}
\label{s.III}

\subsection{\bf Near Neighbors}
Let us start the analysis of the local packing configurations by exploring local neighborhoods.
The average number of spheres in contact with any given sphere is the primary and most investigated parameter in the literature on granular packings \cite{Smith29,Bernal60,Scott62,Mason66,Mason68,Steinhardt83,Seidler00,ppp,Sederman01,Torquato01,Silb02,Kohonen04}.
Indeed, this is a very simple topological quantity which gives information about several important properties of the system.
Unfortunately, although simple in its definition, such a number is an ill-defined quantity from a experimental point of view.  
The reason is that, from a geometrical perspective, the information about the positions, and eventually the  sizes of all spheres  is not sufficient to determine such a number: two spheres can be infinitesimally close but not in touch.
In the literature several physical methods have been used \cite{Smith29,Bernal60,Kohonen04},  but they all encounter  problems essentially associated with the uncertainty in the threshold distance used to define the maximum allowed gap between apparently `touching' spheres.

An exact computation of the number of touching spheres  from a geometry alone is, in general, an impossible task since the result is unavoidably affected by the precision on the sphere centres and the polydispersity of the spheres themselves.  
With the data from X-ray tomography, we can calculate the location of the sphere centers with a precision which is well within 1\% of their diameters.
On the other hand, the beads utilized have a polydisperse bead-diameter distribution  with a standard  deviation around  $\sim 0.02d$.
Therefore, statistically, the large majority of neighbors in contact must stay within a radial distance of $1.02d$.
Table~\ref{t.1} reports the values of the average number of neighbors  ($n_t$) computed in $\mathbf G$ at the four different radial distances: $r = d$, $1.02 d$, $1.05 d$ and $1.1 d$ for the 6 samples A-F.  
We observe values for the average number of neighbors between $n_t(r) \sim$ $2.9$ and $8.9$, and an increasing trend with the packing density.

\subsection{\bf Touching Neighbors}
A more precise estimate for the actual number of spheres in contact can be inferred from the behavior of $n_t(r)$ (shown in Fig.\ref{f.c_cum} as function of the radial distance up to $r=1.1d$).
From Fig.\ref{f.c_cum} one can note that above $r=0.98 d$ the number of neighbors grows very steeply up to a `knee' at about $1.02 d$  where a slower growth takes place.
Such a steep growth in the number of neighbors can only be an effect of  the uncertainty on the positions of the sphere centers and of the spread in the statistical distribution of the distances between spheres in contact (which is a consequence of the polydispersity).
Indeed, in the ideal case when all the exact positions of closely packed, identical,  perfect, spheres are known, one would expects that $n_t(r)$ has a discontinuity at $r = d$ (from zero to the number of neighbors in contact $n_c$) followed by some kind of growth for $r > d$.
For real, polydisperse, non-perfect, spheroidal grains, the distance between elements in contact is not a fixed value but instead it is distributed around an average value.
In Fig.\ref{f.gaussCollapse} it is clearly shown that such a distribution of radial distances  is well mimicked  for $r<d$ by a normal distribution. 
As a consequence, we expect that the steep growth of $n_t(r)$ around $r=d$ is well described with a complementary error function (for $r<d$). 
To such error function we must add  the contribution from the `nearly touching' spheres which is expected to become sizable from $r > d$.
We therefore expect to find an error-function behavior up to $r \sim d$ and then a combined contribution from the error-function and some growing law describing the cumulate number of non-touching neighbors within the distance $r$.
Indeed, we verify that the behavior of $n_t(r)$ for $r \le d$ is very well described by a complementary error function normalized to $n_c$:
\begin{equation}
n_t(r)^{fit} = n_c \frac{1 }{\sqrt{2 \pi \xi^2}}  \int_{-\infty}^r \exp(-\frac{(x-d)^2}{2 \xi^2}) dx  \;\;\;.
\label{ncfit}
\end{equation}
Where the value of the mean ($d$)  is the average sphere diameter which was estimated: $d = 25.00$ voxels (samples B D E F)  and $d=30.81$ voxels (samples A C)  \cite{AsteKioloa}.
On the other hand, the variance $\xi$ can be directly estimated from the data by computing the second moments $\left<(r-d)^2\right>$ for  the radial distances between spheres calculated over half distribution in the region $r < d$:
\begin{equation}
\xi = 2 \sqrt{  \frac{  \sum_{i,j } ( r_{i,j} -d )^2 H(d-r_{i,j}) }{  \sum_{i,j}  H(d-r_{i,j}) }}
\label{xi}
\end{equation}
with $i,j$ indices labeling the sphere centers; the symbol $r_{i,j}$ indicates the distances between the centers of sphere $i$ and $j$; and $H(d-r_{i,j})$ the step function which returns 1 if $r_{i,j}<d$ and 0 if $r_{i,j} \ge d$. 
From Eq.\ref{xi} we retrieve variances $\xi$ in the interval between $0.01d$ and $0.02 d$   (all the values  are reported in Table~\ref{t.1} ).
These values are consistent with the bead-polydispersity and are significantly larger than the estimated uncertainty on the sphere centers.
The only free parameter left in Eq.\ref{ncfit} is the value of $n_c$ which can be now computed by best fitting the agreement between the  data for $n_t(r)$ and the function $n_c^{fit}(r)$ in the region $r<d$.
In Fig.\ref{f.c_cum} it is shown that the function $n_c^{fit}(r)$ fits well the data for $r<d$ by using the values of $n_c$ given in Table~\ref{t.1}. 
At larger distances ($r>d$) near-neighbors not in contact start to contribute to  $n_t(r)$ and the two behaviors split.
We estimate that in the 6 samples A-F there are in average between 5.81 and 6.97 spheres in contact (see Table~\ref{t.1} and Fig.\ref{f.Nc_rho}). 
These numbers fall in the range of reported values: Bernal measures $n_c = 5.5$ at $\rho = 0.6$ and $n_c = 6.4$ at $\rho = 0.62$  \cite{Bernal60};  whereas from the data by  Scott  we have $n_c = 7$ at $\rho = 0.63$ \cite{Scott62}.
Note that this latter data has been recalculated from that reported by  applying the deconvolution method described above.

In Fig.\ref{f.Nc_rho} the values of $n_c$ vs. density for the samples A-F are reported together with those from Bernal and Scott.
Such agreement between these different data is remarkable considering the different experimental protocols, the different preparations of the samples, the different criteria for identifying and counting spheres in contact and the different polydispersity of the spheres. 
As one can see from Fig.\ref{f.Nc_rho}, they all show a clear and consistent  increasing behaviour with the density. 
A similar increasing trend was also found in simulated packings \cite{Clarke93,Yang00}.
This dependence on the packing density has important theoretical implications which are discussed hereafter.

\subsection*{Mechanical Equilibrium}

In a stack of grains at mechanical equilibrium, Newton's equations for the balance of force and torque acting on each grain must be satisfied.
Lagrange and Maxwell \cite{Lagrange788,Maxwell864} have been the first to note that  in these kinds of systems, to achieve stability, the number of degrees of freedom must balance  the number of constraints.
It is straightforward to calculate that the balance between freedom and constraints requires $n_c=6$ in the case of perfectly spherical friction-less spheres, and $n_c = 4$ for more realistic grains (non-spherical with friction) \cite{Edwards99,Micoulaut02}.
 These values of $n_c$  are encouraging.
 Indeed, they predict  values which are rather close to the ones observed experimentally.
 However, it must be noted that this condition for $n_c$ is neither sufficient nor necessary \cite{Moukarzel96,Graver,Thorpe}.
 Indeed, there can be local configurations which contribute to $n_c$ but do not contribute to the rigidity of the whole system.
 (These are, for instance, the `rattling' grains which can be removed from the system without affecting its stability.)
 On the other hand, there are local arrangements which satisfy the counting rule on $n_c$ but nevertheless are not rigid \cite{Moukarzel96,Graver}.

In recent years there have been scores of theoretical approaches which consider real, disordered, granular packings to be isostatic (free of self induced stresses) \cite{Moukarzel98,Edwards99}.
The advantage is that in a system at isostatic equilibrium, the intergranular forces are uniquely determined by the balance of force and torque alone.
On the contrary, a overconstrained structure can generate self-stress and the deformation of individual grains becomes relevant. 
 In real granular materials (or in bead packs) friction and rotational degrees of freedom are unavoidable, therefore the Maxwell counting implies that isostatic configurations must have average connectivity of $n_c = 4$.
Unfortunately, such a value is rather small in comparison with all available empirical estimations.
Moreover,  all the experimental observations conclude that $n_c$ increases with the packing density excluding therefore the possibility to fixing $n_c$ at 4 for all sTable~packings.

However, in related studies for the rigidity in network glasses \cite{Thorpe,Chubynsky01}  it has been observed that there can exist two phase transitions associated with  the increase of connectivity in a network: a \emph{rigidity percolation} and a  \emph{stress percolation}, and between these two thresholds an \emph{intermediate phase} which is rigid but unstressed.
The \emph{rigidity percolation}  occurs at the threshold predicted by the Maxwell counting ($n_c = 4$) and appears to be a second order of transition \cite{Chubynsky01}. 
The  \emph{stress percolation} transition occurs at an higher value of $n_c$ and it could be a first-order transition  \cite{Chubynsky01}.
This suggests that  granular packings might be in a \emph{marginal state} between the rigidity and the stress percolation thresholds.
In such an isostatic unstressed state, the system has zero elastic modulus (in the thermodynamic limit) \cite{Thorpe,Chubynsky01},  it is therefore marginally rigid and it can be seen as in a intermediate state between fluid and solid  \cite{Ball02}.
An extrapolation from the experimental data for $n_c$, reported in Fig.\ref{f.Nc_rho}, suggests that  a grain connectivity equal to 4 could  be reached by the system at the density $\rho = 0.55$. 
This would place the rigidity percolation threshold at the loose packing limit \cite{Onoda90}.

%\vspace*{3cm}
\begin{figure}
\begin{center}
\resizebox{0.30\textwidth}{!}{%
%\begin{tabular}{r}
\includegraphics{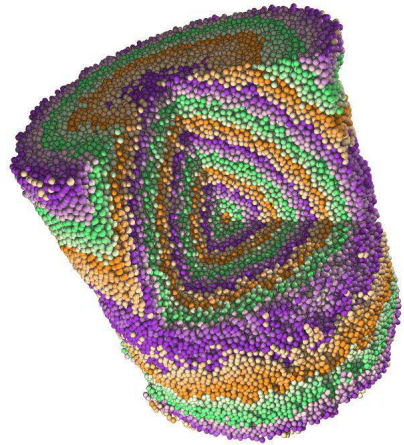} 
%\end{tabular}
}
\end{center}
\caption{ Same as Fig.\ref{f.pack} with a  portion removed and the topological distances from a given central sphere highlighted in colours (online
version).  }
\label{f.pack1}
\end{figure}

%\vspace*{3cm}
\begin{figure}
\begin{center}
\resizebox{0.50\textwidth}{!}{%
\includegraphics{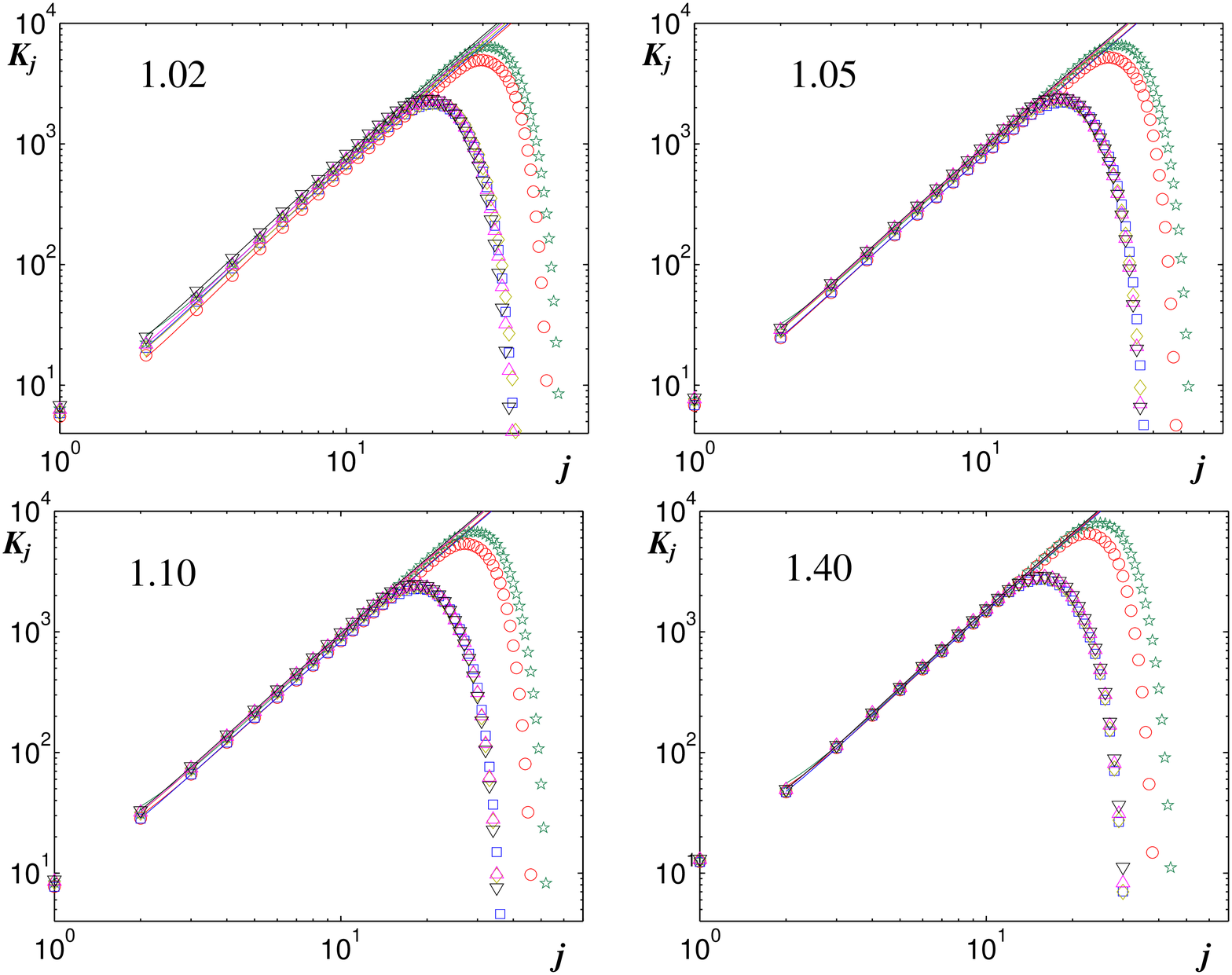} 
}
\end{center}
\caption{ Shell occupation numbers vs. topological distance. 
The symbols indicate the different samples (as in Fig.\ref{f.c_cum}) and the lines are the best-fits (of the growing part only) using the polynomial form: $K_j = a  j^2 + c_1 j + c_0$.
The fits are between $j=2$ and $\hat j =$ 10 (for samples B, D, E, F ) and $\hat j =$ 15 (for samples A, C).
The data refer to threshold distances $1.02d$, $1.05d$, $1.1d$ and $1.4d$ as indicated in the figures. 
 }
\label{f.shell}
\end{figure}

%\vspace*{3cm}
\begin{figure}
\begin{center}
\resizebox{0.40\textwidth}{!}{%
\includegraphics{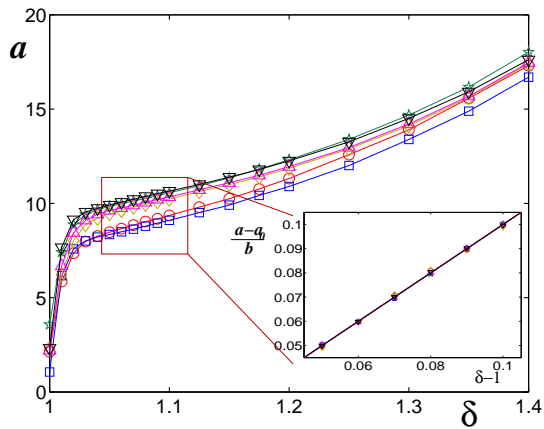} 
}
\end{center}
\caption{ The coefficient $a$ increases with the threshold $\delta$ on the radial distance ($\delta$ is expressed in sphere-diameters unit).
The insert shows that, for all the samples,   the coefficient $a$ follows the linear law $a \sim b(r-d) + a_0$  in the region between $1.05$ and $1.1$.
}
\label{f.shell1}
\end{figure}

\section{ Contact Network:  Beyond First Neighbors}
\label{s.CNBFN}
\label{s.IV}

Any force-path or any infinitesimal local grain displacement must mechanically propagate from grain to grain through the network of touching grains.
The understanding of the hierarchical organization of such contact network beyond first neighbors is therefore of great importance.
Here we apply to granular matter an approach which was originally developed for the study of crystalline systems \cite{Brunner71,Brunner79,OKeeffe91,OKeeffe91b,OKeeffe95,Conway97} and disordered foams \cite{ASBORI,asteSoap96,Aste99shell}.
The topological structure of crystalline frameworks has been intensely studied in terms of the number of atoms that are $j$ bonds away from a given atom \cite{Brunner71,Brunner79,OKeeffe91,OKeeffe91b,OKeeffe95,Conway97}.
If we start from a given `central' atom, the first `shell' (distance $j=1$) is made by all the atoms in contact with the central one.
The second shell (distance $j=2$) consists of all atoms which are neighbors to the atoms in the first shell, excluding the central one.
Moving outward, the  atoms at shell $j+1$  are all the ones which are bonded to atoms in shell $j$ and which have not been counted previously.
In infinite, periodic, crystalline structures with no boundaries,  the number of atoms per shell should increase with the topological distance and it has been shown that in several three-dimensional crystalline structures  the law of growth for the number of atoms ($K_j$) at shell $j$ can be described with: $K_j = a_j j^2 + b_j j + c_j$  (with $a_j$,  $b_j$ and $c_j$ coefficients that might vary with $j$ but only within a bounded finite interval) \cite{Brunner71,Brunner79,OKeeffe91,OKeeffe91b,OKeeffe95,Conway97,Grosse96}.
Following the definition of O'Keeffe \cite{OKeeffe91b}, for these crystalline systems, the asymptotic behavior  of $K_j$ can be characterized in terms of an `exact topological density' : $TD = \left< a_i \right>/3$ \cite{Grosse96}.
It has been noted that such a  topological density is interestingly related to the geometrical density of the corresponding crystalline structure and it is a powerful instrument to characterize such systems.
For instance, it is easy to compute that the cubic lattice has $K_j = 4 j^2 +2$.
Whereas, spheres packed in a $bcc$ (body centered cubic) crystalline arrangement have: $K_j = 6 j^2 +2$ ($j>0$). 
On the other hand, it has been shown \cite{Conway97} that for Barlow packings of spheres,  $K_j$ are always in a narrow range within:
\begin{equation}
\label{barlow}
10 j^2 + 2 \le K_j \le \lfloor \frac{21 j^2 }{2} \rfloor + 2 \;\;\;\; (j > 0)\;\;,
\end{equation}
where the brackets $ \lfloor ...  \rfloor$ indicate the floor function.
In Eq.\ref{barlow} the lower limit is associated with the $fcc$ (face centered cubic) packing and the upper limit corresponds to the  and $hcp$ (hexagonal closed packed)  packing. 
It has been observed by O'Keeffe and Hyde \cite{OKeffeBook} that for  \emph{lattice sphere packings}  the general rule holds:  $K_j = (n_c -2) j^2  + 2$, implying therefore $a = n_c -2$.  %Chaper 6 pag 270

%\vspace*{3cm}
\begin{figure}
\begin{center}
\resizebox{0.40\textwidth}{!}{%
\includegraphics{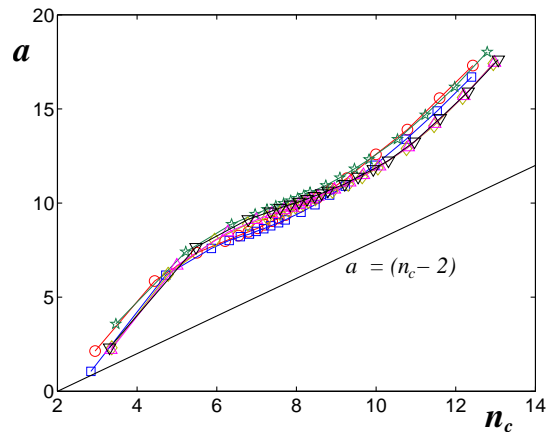} 
}
\end{center}
\caption{ The coefficient $a$ plotted against the average number of neighbors in contact $n_c$ show that disordered packings have larger topological densities in comparison with \emph{lattice} sphere packings. 
}
\label{f.shell2}
\end{figure}

Beyond perfect crystalline order very few results are known either from theoretical, empirical or numerical point of view.
One can argue that $K_j$ must grow with a law comparable with the law for a spherical shell: $K_j \sim a j^2 \sim 4 \pi j^2$.
However, it is also clear that the shape of the growing shell and its roughness can drastically change the coefficient $a$ (as observed in two dimensional cases \cite{asteSoap96,aste2Dfroth}).
Moreover, it can be shown \cite{ADHhypnet04} that in some topological networks the law of growth can follow an intrinsic dimension which is different from the dimension of the embedding space (3 in our case).
This mechanism can produce power law growth with exponents different from 2, or different behaviors such as exponential -- or even faster --  laws of growth  \cite{ADHhypnet04}.

We observed that the number of spheres at a given topological distance $j$ from a central one follows a power law growth (see Fig.\ref{f.shell}) until a critical distance $\hat j$, above which the shells hit the sample boundaries and $K_j$ starts to decrease.  
We verify that a quadratic law $K_j = a j^2 + c_1 j + c_0$ fits quite accurately  the observed behaviors of $K_j$ for $j < \hat j$.
This fixes the intrinsic dimension for these systems equal to 3 (which coincides with the geometrical dimension of the embedding space).
The coefficient $a$ depends on  the threshold $\delta$ on the radial distances within which we consider spheres to be connected in the contact network.
Indeed, changes in the threshold distances are unavoidably associated with changes in the contact network and an enlargement of the threshold distance must correspond to a thickening of the shell. 
In Fig.\ref{f.shell1} we show that the dependence of the coefficient $a$ on the threshold distance $\delta$ is rather complex and not  reducible to a simple law.
However, we verify that in the interval of threshold distances  between $1.05d$ and $1.1d$, a simple linear increment is observed: $a \simeq b\delta + a_0$ (typically with $b \sim O(1)$).
This is shown in the insert in Fig.\ref{f.shell1}.
Such a law suggests that a unique value for the topological density, independent from the threshold distance, can be associated with the value of the coefficient $a$ at $\delta = 0$: $T_D = a_0/3$.
In Table~\ref{t.2} are reported the  topological densities  for all the samples A-F.
As one can see the quantity $3 T_D = a_0$  stays in a rather narrow range around 8.5 and slightly increases with the sample density.
Interestingly we observe in Fig.\ref{f.shell2} that disordered sphere packings have coefficient $a$ consistently above   $n_c-2$ implying therefore that such packings have larger topological densities than lattice sphere packings with the same coordination number.
This observation might be relevant when the structural stability and rigidity of such system is concerned.
A view of the topological shell structure constructed from a given central sphere in one of the samples, is shown in Fig.\ref{f.pack1}.

%It is worth noting that in these sphere packings the law of grow for the shell occupation numbers appear to be little dependent on the the sample density especially when large threshold distances are considered (Figs.\ref{f.shell} and \ref{f.shell1}).

\begin{table*}
\begin{tabular}[c]{ccccccc}
\hline
&  $3 T_D$ & $r$ treshold  & $(\hat Q_4,\hat Q_6)$ & $dis$ (\%) & $fcc$ (\%)& $hcp$   (\%) \\
\hline
A  & $ 7.2 \pm 0.3$ &
 \begin{tabular}[c]{c} $1.1$\\ $1.2$\\ 1.3 \\ 1.4 \end{tabular} &
 \begin{tabular}[c]{c} $(0.27,0.47)$ \\ $(0.22,0.42)$\\ $(0.18,0.40)$ \\ $(0.15,0.36)$ \end{tabular}  &
 \begin{tabular}[c]{c} $23$\\ $32$\\ 38\\ 42 \end{tabular} &
 \begin{tabular}[c]{c} $3$ \\ $2$ \\ $1$ \\ 2 \end{tabular} &
 \begin{tabular}[c]{c} $1$ \\ $3$ \\ $5$ \\ 4 \end{tabular}   \\ \hline 
B  & $ 7.2 \pm 0.4$&  
  \begin{tabular}[c]{c} $1.1$\\ $1.2$\\ 1.3\\ 1.4 \end{tabular} &
 \begin{tabular}[c]{c} $(0.30,0.45)$ \\ $(0.23,0.44)$\\ $(0.16,0.38)$ \\ $(0.14, 0.35)$ \end{tabular}  &
 \begin{tabular}[c]{c} $24$\\ $32$\\ 37\\ 43 \end{tabular} &
 \begin{tabular}[c]{c} $3$ \\ $2$ \\ $1$ \\  2 \end{tabular} &
 \begin{tabular}[c]{c} $1$ \\ $3$ \\ $5$ \\ 5 \end{tabular}  \\  \hline 
C & $  8.7 \pm 0.4$ & 
  \begin{tabular}[c]{c} $1.1$\\ $1.2$\\ 1.3\\ 1.4 \end{tabular} &
  \begin{tabular}[c]{c} $(0.23,0.46)$ \\ $(0.21,0.43)$\\ $(0.15,0.40)$ \\ $(0.12,0.37)$ \end{tabular}  &
 \begin{tabular}[c]{c} $28$\\ $35$\\ 41 \\ 45 \end{tabular} &
 \begin{tabular}[c]{c} $5$ \\ $3$ \\ $1$ \\ 3 \end{tabular} &
 \begin{tabular}[c]{c} $2$ \\ $7$ \\ $11$ \\ 8 \end{tabular}  \\  \hline 
D & $  8.4 \pm 0.3$ & 
  \begin{tabular}[c]{c} $1.1$\\ $1.2$\\ 1.3\\ 1.4 \end{tabular} &
  \begin{tabular}[c]{c} $(0.25,0.44)$ \\ $(0.19,0.44)$\\ $(0.15,0.40)$ \\ $(0.11,0.36)$ \end{tabular}  &
 \begin{tabular}[c]{c} $28$\\ $35$\\ 42 \\ 46 \end{tabular} &
 \begin{tabular}[c]{c} $4$ \\ $2$ \\ $1$ \\ 1 \end{tabular} &
 \begin{tabular}[c]{c} $1$ \\ $7$ \\ $11$ \\ 8 \end{tabular}  \\  \hline 
E  & $ 8.6 \pm 0.4$ &  
  \begin{tabular}[c]{c} $1.1$\\ $1.2$\\ 1.3\\ 1.4 \end{tabular} &
  \begin{tabular}[c]{c} $(0.22,0.44)$ \\ $(0.20,0.43)$\\ $(0.15,0.39)$ \\ $(0.12,0.36))$\end{tabular}  &
 \begin{tabular}[c]{c} $27$\\ $37$\\ 42 \\ 47 \end{tabular} &
 \begin{tabular}[c]{c} $5$ \\ $3$ \\ $1$ \\ 2 \end{tabular} &
 \begin{tabular}[c]{c} $2$ \\ $7$ \\ $12$ \\ 10 \end{tabular}  \\  \hline 
F  & $ 8.9 \pm 0.4$ & 
  \begin{tabular}[c]{c} $1.1$\\ $1.2$\\ 1.3\\ 1.4 \end{tabular} &
  \begin{tabular}[c]{c} $(0.23,0.44)$ \\ $(0.16,0.45)$\\ $(0.13,0.42)$ \\ $(0.10,0.38)$ \end{tabular}  &
 \begin{tabular}[c]{c} $31$\\ $38$\\ 43 \\ 47 \end{tabular} &
 \begin{tabular}[c]{c} $6$ \\ $4$ \\ $1$ \\ 3 \end{tabular} &
 \begin{tabular}[c]{c} $4$ \\ $12$ \\ $17$ \\ 13 \end{tabular}   \\
\hline
\end{tabular} 
\caption{ 
\label{t.2} Topological Densities ($3 T_D = a_0$). Most recurrent values for the local orientation order $(\hat Q_4,\hat Q_6)$. Fraction of local configurations with $( Q_4,Q_6)$ in the range  $(\hat Q_4 \pm 0.05 ,\hat Q_6 \pm 0.05 )$ ($dis$).
 Fraction of local configurations close to special form of  order: 
($fcc$)  with $( Q_4, Q_6)$ in the range $(0.191 \pm 0.05 ,0.574 \pm 0.05 )$ ;  
 ($hcp$) with $(Q_4, Q_6)$ in the range $(0.097 \pm 0.05 ,0.485 \pm 0.05 )$ .
}
\end{table*}

\section{ Local Orientation }
\label{s.LO}
\label{s.9}
\label{s.V}

Revealing and quantifying orientational order is a key issue in establishing the nature of internal organization, and in particular in determining whether there exists a `typical' disordered state or identifying possible tendencies towards hidden symmetries.
Indeed, if such a `typical' state  exists or/and if there is a tendency toward a specific local organization, then it will be possible to associate to a given granular pack an order parameter which could measure how close the packing is to the ideal structure. 
On the other hand if one can prove that the system is a collection of uncorrelated local configurations then this will make it possible to calculate the configurational entropy and -consequently- the probability to find the system in a given state at a given density.
It has been often argued that the competition between the tendency to form a locally compact configuration and the geometrical frustration could be the key to understand the mechanism of formation of disordered packings and glassy structures.
If this is the case we will expect to see at local level, configurations with rotational symmetries characteristic of icosahedral and other closed packed structures.
The study of the local rotational symmetry can therefore give  insights also on the mechanism of formation of these structures.  

%\vspace*{3cm}
\begin{figure}
\begin{center}
\resizebox{0.50\textwidth}{!}{%
\includegraphics{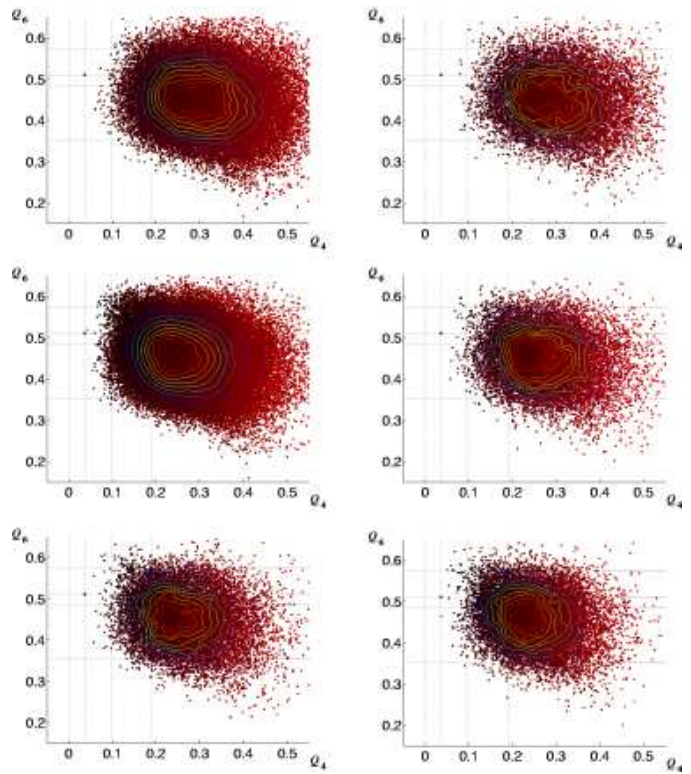} 
}
\end{center}
\caption{ 
Values of $(Q_4,Q_6)$ for all the local configurations in samples `A-F' given a threshold distance of 1.1$d$.
Each dot correspond to a given sphere in $\mathbf G$.
The color (online version) ranges from light-red to dark-red depending on the number of neighbors (within the threshold distance) of each local configuration.
The lines are contour  plots of the frequencies. 
The positions of specific symmetries ({\it ico, sc, bcc, fcc, hcp}) in the  $(Q_4,Q_6)$ plane are also indicated with `*' and projected on the axes with dotted lines .
}
\label{f.QL}
\end{figure}
%\vspace*{3cm}

\begin{figure}
\begin{center}
\resizebox{0.50\textwidth}{!}{%
\includegraphics{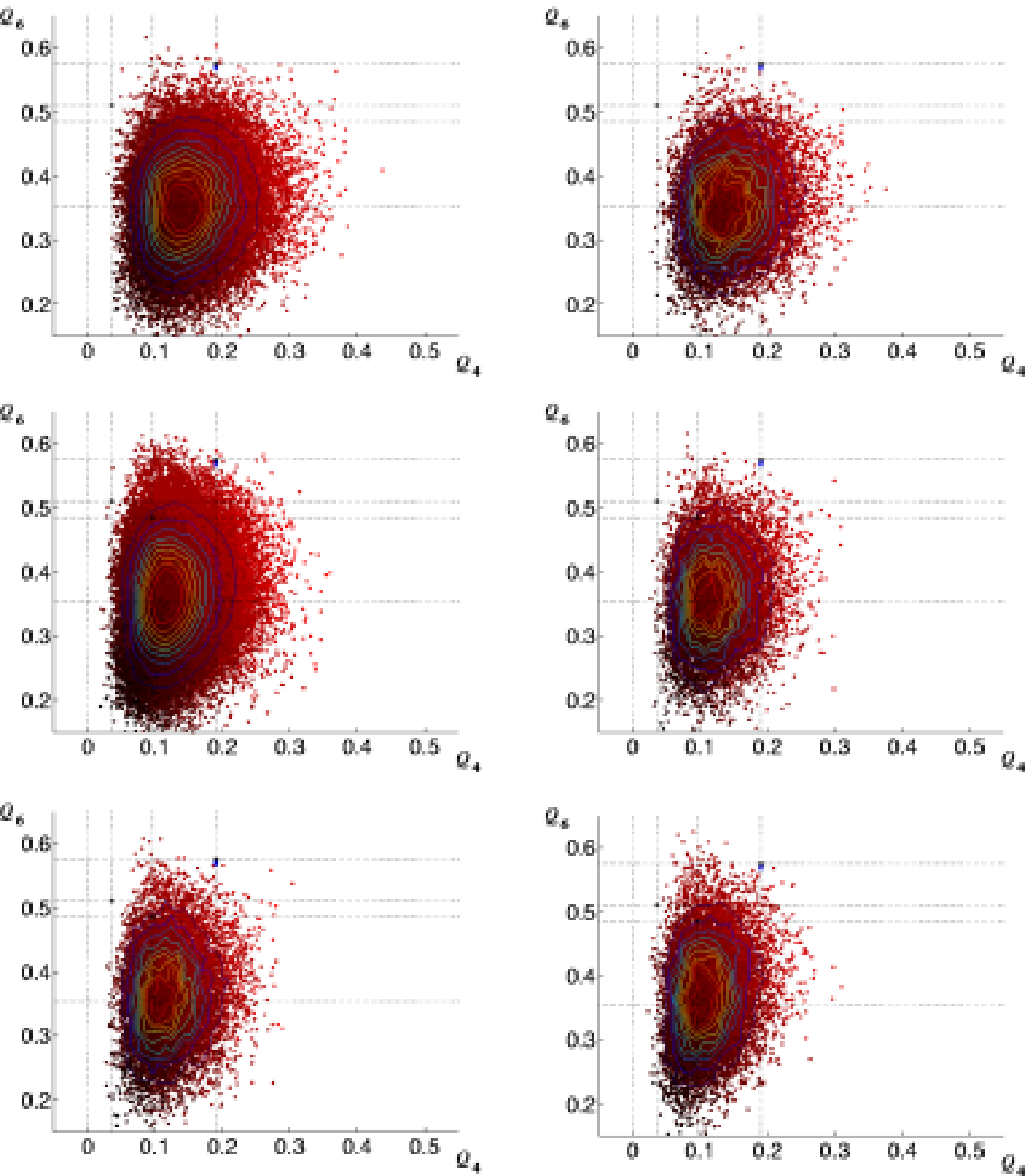} 
}
\end{center}
\caption{ 
Same plot as in Fig.\ref{f.QL} but with threshold distance 1.4$d$.
}
\label{f.QL1}
\end{figure}

The challenge is to find a measure of rotational symmetry which  is  invariant with respect to rotations in the system of coordinates. 
A powerful solution was introduced by Steinhardt, Nelson and Ronchetti \cite{Steinhardt83} by assigning a set of spherical harmonics $Y_{l,m}(\theta(\vec r),\phi(\vec r))$ to the vectors $\vec r$ between couples of spheres  (with $\theta(\vec r)$ and $\phi(\vec r)$ the polar and azimuthal angles of $\vec r$) and introducing the quantities:
\begin{equation}
Q_l = \sqrt{ \frac{4 \pi}{2l+1} \sum^l_{m=-l}  |\left< Y_{l,m}(\theta(\vec r_i),\phi(\vec r_i)) \right>|^2}\;\;,
\label{ee}
\end{equation}
with average $\left<(...)\right>$ over the bonds `$i$' in the sample.
Such a quantity is invariant under rotations in the coordinate system and it takes characteristic values which can be used to quantify the kind and the degree of rotational symmetry in the system.
However, it must be noted that the quantity $Y_{l,m}(\theta(\vec r_i),\phi(\vec r_i))$ depends on the orientation, therefore in the case of a polycrystalline aggregate, with finite correlation length, its average $\left< Y_{l,m}(\theta(\vec r_i),\phi(\vec r_i)) \right>$ will decrease and tend to zero with the sample-size.
To avoid this inconvenience, which makes the comparison between values of $Q_l$ on differently sized samples meaningless, it is convenient to adopt a local measure of $Q_l$ by  restricting  the average only over the local bonds between a sphere and its neighbors.
In this way, to each sphere in the system can be associated  a $Q_l$ and local order can be singled out by counting the number of configurations with $Q_l $ corresponding to special symmetries.
In particular the two cases $l=4$ and $l=6$ have special significance.
For instance, the simple cubic lattice has $(Q_4,Q_6)^{sc} = (0.764, 0.354)$, the body centered cubic lattice has $(Q_4,Q_6)^{bcc} = (0.036, 0.511)$, the $fcc$ has  $(Q_4,Q_6)^{fcc} = (0.191, 0.574)$, the $hcp$ has $(Q_4,Q_6)^{hcp} = (0.097,  0.485)$ and the icosahedral rotational symmetry gives $(Q_4,Q_6)^{ico} = (0, 0.663)$.
Since the lowest  non-zero $Q_l$ common to the icosahedral, hexagonal and the cubic symmetries is for $l=6$, it has been argued by several authors that the value of $Q_6$ is a good indicator of the degree of order in the system and it might be used as an `order parameter'  \cite{RichardPRE99,RichardEPL99,Trusk00,Torquato00,Kansal02}.
Indeed, $Q_6$ is very sensitive to any kind of crystallization and it increases significantly when order appears \cite{RichardPRE99}.

Similarly to that discussed in the previous sections,  the measure depends  on the adopted geometrical criteria to identify neighbors.
In the literature, several different criteria are used:  in \cite{Steinhardt83} all neighbors within $1.2d$ are considered; in \cite{Trusk00} the neighbors up to the radial distance which correspond to the first minimum in the radial distribution function ($r\sim 1.4d$, see section \ref{s.gr}) was considered;  in \cite{RichardPRE99} and \cite{Kansal02} the Vorono\"{\i} (or Delaunay \cite{Voronoi,ppp}) neighbors where used instead.
This last definition might be misleading in some cases (as pointed out by \cite{Steinhardt83}), since the  Vorono\"{\i} method tends to associate bonds to sometime distant neighbors.
For instance,  an $fcc$ crystalline arrangement (with infinitesimal perturbation) takes 2 extra neighbors (from 12 to 14 in average) using the Vorono\"{\i} criteria.
Here the influence of the neighboring criteria is analyzed by using 4 different threshold distances: $1.1d$, $1.2d$, $1.3d$ and $1.4d$.

Examples of the distribution of local $( Q_4, Q_6 )$ are shown in Figs.\ref{f.QL} and \ref{f.QL1}.

We observe values of $(Q_4,Q_6)$  narrowly distributed around their most recurrent values $(\hat Q_4,\hat Q_6)$  with very large fractions of local configurations (between 23 and 47 \%) which have local symmetries characterized by $(Q_4,Q_6)$ within the range $(\hat Q_4 \pm 0.05 , \hat Q_6 \pm 0.05 )$.
The values of  $(\hat Q_4,\hat Q_6)$ range between $0.10 \le \hat Q_4 \le 0.30$ and $0.35 \le \hat Q_6 \le 0.45$ across all samples and all thresholds (see Table~\ref{t.2}).
Such values are  far from any special symmetry.
In order to search for signatures of known local symmetries we measured the fraction of local configurations with $(Q_4,Q_6)$ in a region within a range $\pm 0.05$ from the values in the ideal structures ( \textit{fcc}, \textit{hcp},  $ico$, $sc$, and $bcc$).  
We found that there are no significant fractions  (below 1 \%)  of local configurations with  symmetry compatible with icosahedral, simple cubic or $bcc$; there is a small fraction (between 1 to 6 \%) of configurations with local symmetry compatible with $fcc$, and there is a fraction of configurations with $hcp$-kind of local order which becomes quite significant at large densities  (reaching 17\% at $\rho = 0.64$ and $\delta = 1.3d$).
This occurrence of a rather large fraction of local symmetry with an $hcp$-like character might suggest the beginning of a crystallization process.
However, we have verified that there are no correlations between neighboring sites with symmetry close to $hcp$. 
This excludes the  presence of any long range $hcp$ order recurrent or symmetrical organizations beyond first neighbors.

These findings cast considerable doubts over the existence of any crystalline order and also question the idea that a tendency toward local -frustrated-  icosahedral order can be responsible for the resilience to crystallize for such  packings.
These results will be confirmed and reinforced in Section \ref{s.VIII} where the local densities are studied. 
On the other hand, these findings are not conclusive because the origin and nature of the most abundant configurations with $(Q_4, Q_6)  \sim (0.25,0.45)$ are still elusive.
Further studies to clarify the nature and the origin of such local configurations are needed.

%\vspace*{3cm}
\begin{figure}
\begin{center}
\resizebox{0.50\textwidth}{!}{%
\includegraphics{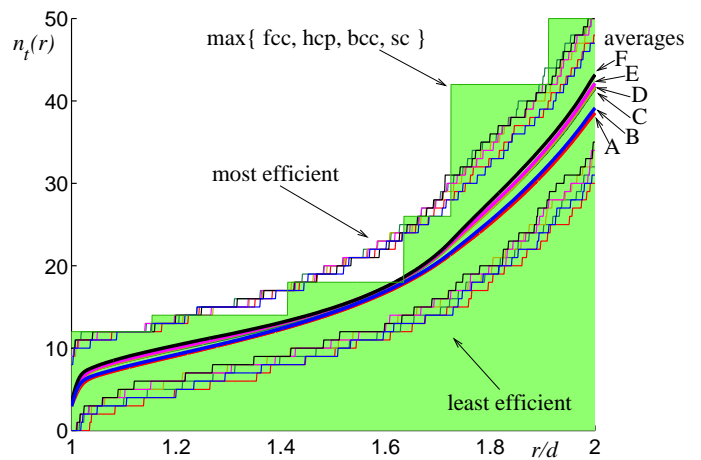} 
}
\end{center}
\caption{ Total number of sphere centers  within a given radial distance $r$: average, maximum and minimum for all the samples.
The filled area is bounded at the top by the most efficient packings among the $fcc$, $hcp$, $bcc$, $sc$.}
\label{f.nt}
\end{figure}

\section{Packing Efficiency}
\label{s.pe}
\label{s.VI}

In this section we investigate how the global sample density is perceived by a sphere at the local level and how the global packing affects the local environment.
To this purpose we compute the number of spheres placed within a certain radial distance from a given sphere.  
This quantity ($n_t(r)$) can be viewed as a measure of how efficiently local dense agglomerates of spheres are formed. We called this measure the \emph{packing efficiency}   \cite{AsteKioloa}.  
It  is well known that no more than 12 spheres can be found in contact with one sphere (the `Kissing number' \cite{ppp}), but the upper limit for the number of spheres within a given radial distance is, in general, unknown.  
Fig.\ref{f.nt} shows the average, the  maximum and the minimum numbers of neighbors within a given radial distance from any sphere in $\mathbf G$.
Clearly,  there are no neighbors up to distances close to $r \sim d$, when suddenly the number of neighbors increases very steeply and then, after this jump,  it  increases with distance following a less steep trend with very comparable behaviors between all 6 samples A-F.
It is of some interest to compare the values of $n_t(r)$ empirically obtained for these disordered samples with the known ones associated with crystalline structures. 
We observe that  in a large range of radial distances between $1d$ and $2d$ there are some local configurations with packing efficiencies which are above the crystalline ones. 
Moreover,  in the region around $r \sim 1.6d$, disordered packings show better \emph{average} packing efficiencies than the crystalline ones ($fcc$, $hcp$, $bcc$, $sc$).  
This is rather surprising if we consider that the $fcc$ and $hcp$ packings are more than 15\% denser than the disordered ones.
However, one can note that increasing the distance, the disordered packings become less and less efficient in comparison with the close packed crystalline arrangements, and above $2d$ all the configurations have a smaller cumulate number of neighbors than the close crystalline packings.
Interestingly, in the same region around $1.6d$ where the disordered packings are very efficient we also observe the minimum spread for the values of the efficiencies across the samples at different densities.
We discuss in further details the behavior of the number of neighbors with the radial distance in the next Section where the radial distribution function is analyzed.

%\vspace*{3cm}
\begin{figure}
\begin{center}
\resizebox{0.40\textwidth}{!}{%
\includegraphics{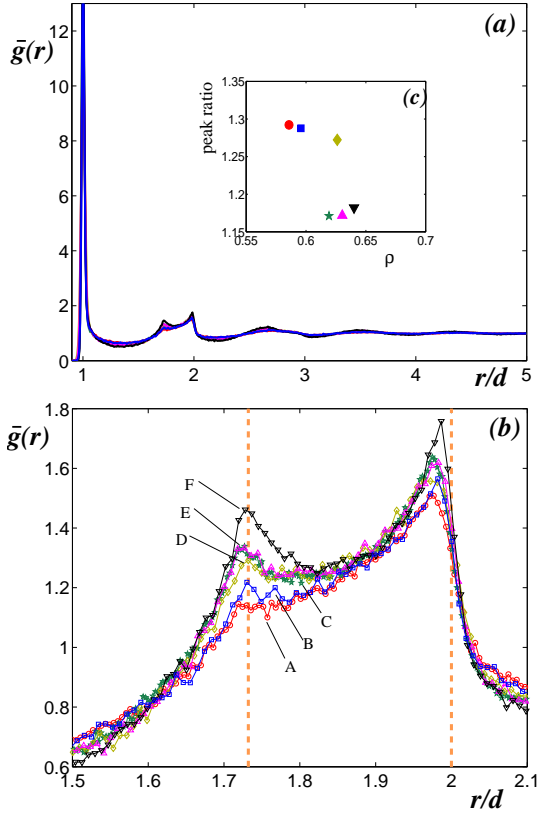} 
}
\end{center}
\caption{ (a) Normalized radial distribution function. (b) The detail of the two peaks respectively at  $\sqrt{2}d$ and $2d$ (vertical lines). (c) The ratio between the value of the peak at $2d$ and the one at $\sqrt{3}d$.}
\label{f.radial}
\end{figure}

%\vspace*{3cm}
\begin{figure}
\begin{center}
\resizebox{0.40\textwidth}{!}{%
\includegraphics{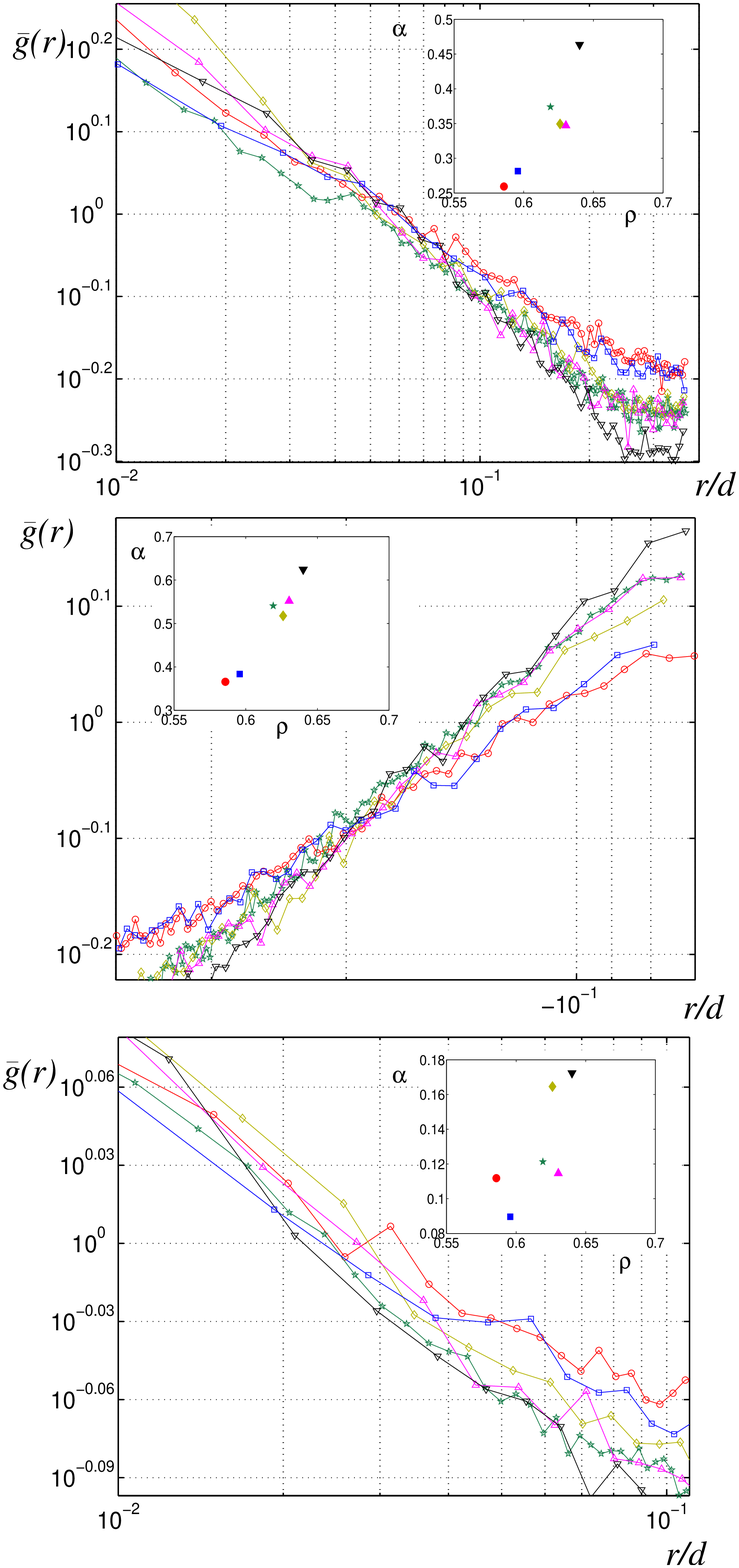} 
}
\end{center}
\caption{ The three peaks of the  Radial distribution function respectively (from top to bottom) at $r=1$, $\sqrt{3}$ and 2 can be described with power law singularities: $g(r) \sim c_0 |r - r_0|^{-\alpha}$.
The coefficient $\alpha$ depends on the sample-densities and their behaviors are reported in the inserts.}
\label{f.exp}
\end{figure}

\section{ Radial Distribution function }
\label{s.PERDF}
\label{s.gr}
\label{s.8}
\label{s.VII}

The Radial Distribution Function ($g(r)$) is the probability distribution to find the \emph{centre} of a particle in a given position at distance $r$ from a reference one.
This measurement is widely used in geometrical characterization of packing structures  and contains information about long range interparticle correlations and their organization \cite{Scott62,Mason68,DGM93}.

In order to calculate this quantity one must count the number of sphere centers within a radial distance $r$ from a given sphere centre.
The average of this number is the quantity  $n_t(r)$ studied in the previous section and it is related to the radial distribution function by:
\begin{equation}
n_t(r_1) - n_t(r_0) =  \int_{r_0}^{r_1} g(r) 4 \pi r^2 dr \;\;\;.  
\label{ntgr}
\end{equation}
Therefore, given the position of the sphere-centers, these two quantities $n_t(r)$ and $g(r)$ can be straightforwardly computed.
Here, we calculate the normalized radial distribution function $\tilde g(r)$ which is  the average number of sphere centres, within a radial distance $r - \Delta /2$ and $r + \Delta /2$, divided by $c r^2$.  
With the constant $c$ fixed by imposing that asymptotically $\tilde g(r)\rightarrow 1$ for $r \rightarrow \infty$.  
(We have verified that different choices of $\Delta$ within a broad range of $10^{-4}d$ to $10^{-2}d$ lead to almost indistinguishable results.)

In Fig. \ref{f.radial} the behavior of  $\tilde g(r)$ vs. $r/d$ is shown.
We observe a very pronounced peak at $r=d$ which corresponds to the neighbors in contact. 
Then the probability to find neighbors decreases with $r$ reaching a minimum around $1.4d$.
Subsequently, at larger radial distances, the probability increases again forming two peaks respectively at $r=\sqrt{3}d $ and $r \simeq 2d$ and then after these peaks it continues to fluctuate with decreasing amplitudes.
The details of the second and third peaks, plotted in Fig.\ref{f.radial}b, show that the two peaks at $r=\sqrt{3}d $ and $r \simeq 2d$  both increase in height with the packing density.
With the peak at $r=2d$ growing faster than the one at $r = \sqrt{3}d$ (see Fig.\ref{f.radial}c).
This might indicate an increasing organisation in the packing structure but, on the other hand, no  signs of crystallisation were detected (see, Section \ref{s.V}, Section \ref{s.VIII} ).

For all the samples investigated, we found that the behavior of $\tilde g(r)$ at radial distances between $r \simeq 1d$ and $r \simeq 1.4d$ (between the first peak and the first minimum) can be quite accurately described in terms of a power law singularity:
\begin{equation}
\tilde g(r) \sim \frac{c_0}{|r - r_0|^\alpha} \;\;\; ,
\label{e.exp}
\end{equation}
with good fits for $r_0 = 1.03$ and  $\alpha$ which increase with the sample-density from 0.27 to 0.45  (Fig.\ref{f.exp}a).
A similar behavior, but with $\alpha = 0.5$ and $r_0 = d$, was reported in \cite{Silb02} for numerical simulations.  
A more recent numerical investigation proposes an exponent $\alpha \sim 0.4$ \cite{Donev04b}.
In Fig.\ref{f.exp}, it is also highlighted the growing trend of  $\alpha$ with the density $\rho$. 

Interestingly, the behavior of $\tilde g(r)$ around the other  two following  peaks (respectively at $ r \sim \sqrt{3}d$ and $\sim 2d$) can be described by using similar power law kind of divergences.
In particular the region $1.4 < r < 1.73$ is well fitted by Eq.\ref{e.exp} with $r_0 = 1.8$ and $\alpha$ between $0.37$ and $0.62$.   
Whereas, the region $2 < r < 2.15$ is well fitted by using $r_0 = 2$ and  $\alpha$ between $0.11$ and $0.17$ (Fig.\ref{f.exp}).   
We must stress that these are qualitative behaviors: a reliable fit with a power law trend must be performed over several orders of magnitude in the $x$ and $y$ scales. 
These linear interpolations in log-log scales must therefore  be considered more as indicative behaviors of qualitative laws more than fits.

The origin and the nature of this power-law like behavior around these peaks is rather puzzling.
Indeed although the presence of such peaks clearly  indicate  some kind of organization in the system, on the other hand other analyses, such as the orientational symmetry discussed in Section \ref{s.V}, exclude the presence of any crystalline or polycrystalline pattern in the samples. 
To better understand this issue one must single out the specific organization of the local configurations which contribute to each peak.
This will be the topic of a future paper.

%\vspace*{3cm}
\begin{figure}
\begin{center}
\resizebox{0.40\textwidth}{!}{%
\includegraphics{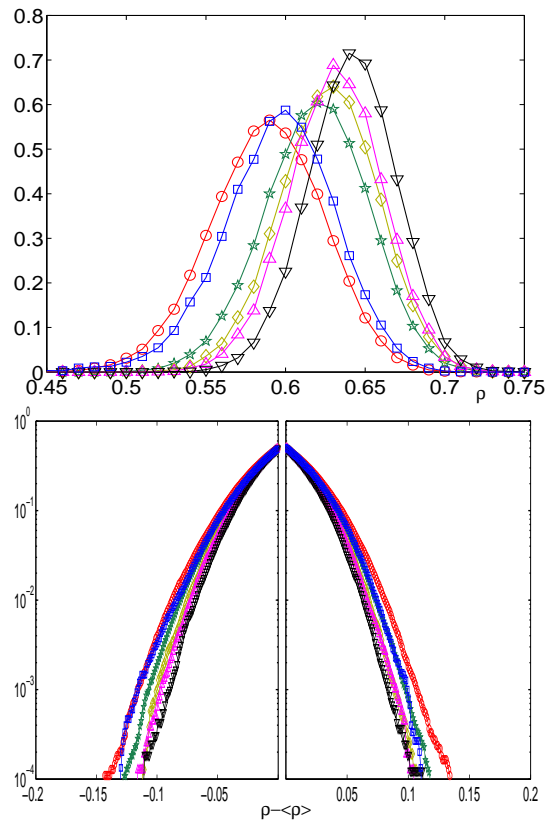}
}
\end{center}
\caption{ 
(above) Distribution of the local densities in $\mathbf G$ for the 6 sapmles.
(below) The cumulants show tails that decrease exponentially or faster with slightly asymmetric distributions in the left and right parts.
}
\label{f.VDens}
\end{figure}

\section{ Global and Local Densities} 
\label{s.v}
\label{s.VIII}

We have already referred previously to the \emph{sample density} which is the fraction between the volume occupied by the spheres divided by the total volume.
Although, this definition is very straightfoward, on the other hand it is well defined only for an infinite sample.
In all the other practical cases, where boundaries are present, the density is unavoidably associated to the way of partitioning space.
A convenient way to study the density is by partitioning the space in local portions and introducing a \emph{local-density} associated with the fraction  of volume occupied within each local portion of space.
Surprisingly, in the literature of granular matter, very few investigations have been devoted to the study of local densities either in experiments or in simulations.
On the contrary the understanding of how the space is shared among the packed spheres and finding how efficiently the spheres can pack locally is an essential information which can contribute to the understanding of both the structure and the formation of these systems.

We calculate the \emph{local-densities} which are defined as the fractions between the sphere-volumes and the volumes of the Vorono\"{\i} cells \cite{Voronoi} constructed around the centre of each sphere in the sample (recall that the Vorono\"{\i} cell is the portion of space closest to a given centre in respect of any other centre).  
The sample-densities are fractions between the sum over the volumes of the spheres in $\mathbf G$ and the sum over the volumes of the Vorono\"{\i} cells associated with these spheres.  
We observe that typically the density is not homogeneously distributed in different parts of the samples. 
This is also discussed by several other works (see for instance \cite{Nowak98,Philippe02}).
In the samples A-F  the  densities are relatively smaller than the average in a region close to the cylinder central axis; the density increases moving outwards from the centre, then it saturates to rather homogeneous values up to a distance of a few (2-3) sphere diameters from the boundary.  
Rather inhomogeneous densities are also observed in the vertical direction, but in this case we find different behaviours depending on the sample-preparation.  
However, we verify that in all the samples sub-regions the densities stay in a rather narrow range (within $\pm$ 0.01, see Table~\ref{t.1}) form the average ones.
More importantly we verify that all the computed structural properties do not change significantly in their behaviors and characteristics with the part of sample analyzed.
In the second column of Table~\ref{t.1}, the average density values for samples A-F  and their interval of variations are reported. 

\subsection*{Local Density distributions and geometrical frustration}

Figure \ref{f.VDens} shows the distribution of the local densities in $\mathbf G$ for the 6 samples.
We observe that these local densities have slightly asymmetric distributions with exponential -or faster decays from the average densities (which are in the range $0.586 \le \rho \le 0.640$) and have standard deviations $\sigma$ within 1.5 \%. 

It has been often argued that the driving mechanism which generates amorphous structures could be the \emph{geometrical frustration}. This derives from the fact that the locally densest configuration in equal sphere packing is achieved by placing on the vertices of  an icosahedron 12 spheres in touch with a central sphere.
But, iocosahedral symmetry  is incompatible with translational symmetry and this generates frustration: some gaps must be formed and the symmetry must be broken.
Indeed, it is known \cite{ppp} that although there exist locally denser configurations, at global scale the densest achievable packings have density $\rho \le \pi/\sqrt{18} = 0.74048...$ which is the one of $fcc$ and $hcp$ crystalline packings.

Here we test whether such a frustration mechanism is really relevant in our granular packs.
To this end we search for  local configurations which are locally close packed at densities larger than the crystalline ones.
If a sizable amount of such configurations is found this implies that indeed the geometrical frustration must have an important role in amorphous systems.
The result is unexpected: only 14 local configurations with densities above 0.7405 were found over a set of more than 209,000 local configurations in the 6 different samples.
Moreover, very few local configurations have density between 0.7 and 0.7405.
Respectively we find: less than 1.7 \% in the densest sample (F at $\rho = 0.640$); less than 1\% in the three samples $C$, $D$, $E$ (with $\rho = 0.617$, $0.630$ and $0.626$); and less than 0.07\% for A and B ($\rho = 0.586$  and  $0.593$)
This is a very strong indication that local sphere-arrangements with high local densities -such as the icosahedral packing -  \emph{play no role} in these disordered sphere packings.

%%%%%%%%%%%%%%%%%%%%%%%%%%%%%%%%%%%%%
%%%%%%%%%%%%%%%%%%%%%%%%%%%%%%%%%%%%
\begin{figure}
\begin{center}
\resizebox{0.40\textwidth}{!}{%
%\begin{tabular}{r}
\includegraphics{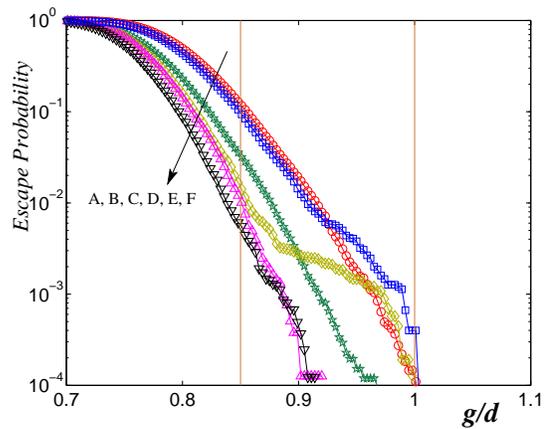} 
%\end{tabular}
}
\end{center}
%\vspace{-0.4cm}
\caption{ 
\label{f.Pe}
Probability to find a path through the first neighbors with section larger than $g$.
}
\end{figure}
%%%%%%%%%%%%%%%%%%%%%%%%%%%%%%%%%%%%%
%%%%%%%%%%%%%%%%%%%%%%%%%%%%%%%%%%%%%

\section{ Geometry and Structural Arrest}
\label{s.IX}

We calculated the gaps between neighboring spheres and evaluated the size of the largest gap for each local configuration  surrounding a given sphere. 
This gives the probability for each sphere to move outward from a given local configuration without displacing the positions of its neighbors.
Such an `escape probability' is computed by constructing circles through the centres of the Delaunay simplexes incident on the central sphere. 
The central sphere can `escape' from this configuration only if its diameter ($d$) is  smaller or equal to the largest radius of such circles. 
Clearly, when such a move is possible, the system can change its geometrical configuration by means of local moves only.
From a thermodynamical point of view this implies that it can dynamically explore the phase-space with low energy moves and reach equilibrium in short relaxation times.
On the contrary, when the escape probability is zero, a structural rearrangement requires the displacement of a larger number of spheres and the system is more likely to be trapped for long times in meta-stable states. 

The relative number of local configurations with gaps larger or equal to a threshold size $g$  is reported in Fig.\ref{f.Pe} for the 6 samples A-F.
We find that all the samples with $\rho > 0.6$ ($C$-$F$) do not have any configuration which allows the central sphere to `escape'.
Whereas samples $A$ and $B$ have few local configurations with gaps larger than $d$ but seems statistically irrelevant ($ < 0.1 \%$).
This strongly suggests that around $\rho \sim 0.58-0.6$ an important phase in the system dynamics reaches an end: above this density, local readjustments involving only the displacement of a single sphere are forbidden.
The particle mobility is constrained mostly within the Vorono\"{\i} cell  and the system compaction can proceed only by involving the collective and correlated readjustment of a larger set of spheres.
At this stage the system can no longer sample the whole phase-space and it is trapped within the basin of attraction of some inherent configuration \cite{AsCo04} and eventually will reach a structural arrest before  the thermodynamical equilibrium is reached.

\section{Conclusions}
The structure of disordered packing of mono-sized spheres has been investigated by means of X-ray Computed Tomography. 
We performed an extensive study over 6 large samples at packing densities ranging between 0.586 and 0.640, investigating several geometrical and topological properties.

The number of neighbors surrounding each sphere in the packing  was studied with unprecedented statistical accuracy (Section  \ref{s.III}).
The average number of spheres in contact was extracted by means of an innovative method which deconvolutes the contribution of touching neighbors  from the contribution from near neighbors. 
The results show that the average number of spheres in contact increases with the sample density and  it is between 5.5 and 7.5 in the range of densities examined (see Fig.\ref{f.Nc_rho}).
An extrapolation to the random loose packing density ($\rho = 0.55$) suggests that at this density the system could have an average number of 4 neighbors per sphere.
This implies the possibility of a rigidity percolation transition taking place at the random loose packing limit.

The structure beyond first neighbors, studied by means of a topological map (Section \ref{s.IV}), shows  that the contact networks of these systems have  intrinsic dimension 3 and topological densities between 7.2 and 8.9 (see Table~\ref{t.2}).
This is a novel approach for three-dimensional non-cystalline structures.
Surprizingly, we found that the topological density in disordered sphere packings is always larger than the topological density in the corresponding lattice sphere packings.
Such a larger topological density is an indication that the contact network is more compact in disordered systems despite the fact that the \emph{geometrical}  density is lower.
This fact might have important implication on the system stability and resilience to perturbations and shocks.

A non-intuitive property was found  by computing the \emph{packing efficency} (Section \ref{s.VI}).
It results that disordered packings can have a larger number of neighbors within a given radial distance than the  crystalline packings (see Fig.\ref{f.nt}).
This fact is surprising if one considers that $fcc$ and $hcp$ packings are more than 15\% denser than the disordered packings.

In Section \ref{s.V} we searched for local symmetries and the existence of significant repetitive local configurations.
The result was that most of the spheres are arranged locally in configurations which are significantly different from any crystalline arrangement excluding therefore the presence of any partial crystallization in these samples.
We also established that there are no statistically significant configurations with icosahedral symmetry.
Moreover in Section \ref{s.VIII} we analyze the local densities of more than 200,000 configurations concluding that there are no statistically significant  local arrangements with density equal or above 0.74.
This excludes that geometrical frustration can play any significant role in the formation of such amorphous packings.

The structural organization emerging from the radial distribution function was discussed in Section \ref{s.VII} where we pointed out  peculiar power law kind of behavior around the peaks of $\tilde g(r)$ at $r=d$, $r=\sqrt{2}d$ and $r= 2d$ (see Eq.\ref{e.exp} and Fig.\ref{f.exp}).
%This is the first time that such behaviors are observed in experiments.

Insights on the dynamical formation of these systems were given in Section \ref{s.IX} where we pointed out that the distribution of gaps between neighboring spheres suggests that a dynamical glass transition might take place at densities around 0.58-0.60.

With the largely increased statistical confidence, these datasets may open new and more fruitful paths to the understanding of granular materials.

\subsection*{Acknowledgements}
The authors gratefully thank Ajay Limaye for the preparations of the volume rendered images.
We thank A. Sakellariou for the help with the reconstruction of the tomographic data. 
We acknowledge several discussion with Alexandre Kabla.
Senden gratefully acknowledges the ARC for his Fellowship.
T. Aste wish to tank M. O'Keeffe for a very useful discussion.
This work was partially supported by the ARC discovery project DP0450292 and Australian Partnership for Advanced Computing National Facilities (APAC).

% \appendix

%\bibliography{beads4}   % Produces the bibliography via BibTeX.

\end{document}